# Using core-shell metamaterial engineering to triple the critical temperature of a superconductor

Vera N. Smolyaninova [1)],  Kathryn Zander [1)],  Thomas Gresock [1)],  Christopher Jensen [1)], Joseph C. Prestigiacomo [2)] , M. S. Osofsky [2)], and Igor I. Smolyaninov [3)]

*[1]Department of Physics Astronomy and Geosciences, Towson University,*

*8000 York Rd., Towson, MD 21252, USA*

*[2] Naval Research Laboratory, Washington, DC 20375, USA*

*[3] Department of Electrical and Computer Engineering, University of Maryland, College Park, MD 20742, USA*

**Recent experiments have shown the viability of the metamaterial approach to dielectric response engineering for moderately enhancing the transition temperature, $T_c$, of a superconductor. In this report, we demonstrate the use of $Al_2O_3$-coated aluminium nanoparticles to form the recently proposed epsilon near zero (ENZ) core-shell metamaterial superconductor with a $T_c$ that is three times that of pure aluminium. IR reflectivity measurements confirm the predicted metamaterial modification of the dielectric function thus demonstrating the efficacy of the ENZ metamaterial approach to $T_c$ engineering. These results provide an explanation for the long known, but not understood, enhancement of the $T_c$ of granular aluminum films.**

Recent theoretical [1,2] and experimental [3] work have conclusively demonstrated that the metamaterial approach to dielectric response engineering can be used to increase the critical temperature of a composite superconductor-dielectric metamaterial. Indeed, according to Kirzhnits *et al.* [4] the superconducting properties of a material may be expressed via its effective dielectric response function, as long as the material may be considered as a homogeneous medium on the spatial scales below the superconducting coherence length. The electron-electron interaction in a superconductor may be expressed in the form of an effective Coulomb potential

$$V(\vec{q},\omega) = \frac{4\pi e^2}{q^2 \varepsilon_{eff}(\vec{q},\omega)}, \qquad (1)$$

where $V = 4\pi e^2/q^2$ is the Fourier-transformed Coulomb potential in vacuum, and $\varepsilon_{eff}(q,\omega)$ is the linear dielectric response function of the superconductor treated as an effective medium. Based on this approach, Kirzhnits *et al.* have derived expressions for the superconducting gap, $\Delta$, critical temperature, $T_c$, and other important parameters of the superconductor. Following this "macroscopic electrodynamics" formalism, it appears natural to use recently developed plasmonics [5] and electromagnetic metamaterial [6] tools to engineer and maximize the electron pairing interaction (1) in an artificial "metamaterial superconductor" [1,2] via deliberate engineering of its dielectric response function $\varepsilon_{eff}(q,\omega)$. For example, considerable enhancement of attractive electron-electron interaction may be expected in such actively studied metamaterial scenarios as epsilon near zero (ENZ) [7] and hyperbolic metamaterials [8]. In both cases $\varepsilon_{eff}(q,\omega)$ may become small and negative in substantial portions of the relevant four-momentum $(q,\omega)$ space, leading to enhancement of the electron pairing interaction. This approach has been verified in experiments with compressed mixtures of tin and barium titanate nanoparticles of varying composition [3]. An increase of the critical temperature of the order of $\Delta T_c \sim 0.15$ K compared to bulk tin has been observed for 40% volume fraction of barium titanate nanoparticles, which corresponds to ENZ conditions. Similar results were also obtained with compressed mixtures of tin and strontium titanate nanoparticles.

These results clearly demonstrated a deep connection between the fields of superconductivity and electromagnetic metamaterials. However, despite this initial success, the observed critical temperature increase was modest. It was argued in [2] that the random nanoparticle mixture geometry may not be ideal because simple mixing of superconductor and dielectric nanoparticles results in substantial spatial variations of $\varepsilon_{eff}(q,\omega)$ throughout a metamaterial sample. Such variations lead to considerable broadening and suppression of the superconducting transition.

To overcome this issue, it was suggested that an ENZ plasmonic core-shell metamaterial geometry, which has been developed to achieve partial cloaking of macroscopic objects [9], should be implemented [2]. The cloaking effect relies on mutual cancellation of scattering by the dielectric core (having $\varepsilon_d>0$) and plasmonic shell (with $\varepsilon_m<0$) of the nanoparticle, so that the effective dielectric constant of the nanoparticle becomes very small and close to that of vacuum (a plasmonic core with a dielectric shell may also be used). This approach may be naturally extended to the core-shell nanoparticles having negative ENZ behaviour, as required in the superconducting application. Synthesis of such individual ENZ core-shell nanostructures followed by nanoparticle self-assembly into a bulk ENZ metamaterial (as shown in Fig.1) appears to be a viable way to fabricate an extremely homogeneous metamaterial superconductor.

The design of an individual core-shell nanoparticle is based on the fact that scattering of an electromagnetic field by a sub-wavelength object is dominated by its electric dipolar contribution, which is defined by the integral sum of its volume polarization [9]. A material with $\varepsilon >1$ has a positive electric polarizability, while a material with $\varepsilon <1$ has a negative electric polarizability (since the local electric polarization vector, $P=(\varepsilon -1)E/4\pi$, is opposite to $E$). As a result, the presence of a plasmonic shell (core) cancels the scattering produced by the dielectric core (shell), thus providing a cloaking effect. Similar consideration for the negative ENZ case leads to the following condition for the core-shell geometry:

$$r_c^3 \varepsilon_c \approx -\left(r_s^3 - r_c^3\right)\varepsilon_s, \qquad (2)$$

where $r_c$ and $r_s$ are the radii, and $\varepsilon_c$ and $\varepsilon_s$ are the dielectric permittivities of the core and shell, respectively. Eq.(2) corresponds to the average dielectric permittivity of the core-shell nanoparticle being approximately equal to zero. Working on the negative side of this equality will ensure negative ENZ character of each core-shell nanoparticle. A dense assembly of such core-shell nanoparticles will form a medium that will have small negative dielectric permittivity. Moreover, in addition to obvious advantage in homogeneity, a core-shell based metamaterial superconductor design enables tuning of the spatial dispersion of the effective dielectric permittivity $\varepsilon_{eff}(q,\omega)$ of the metamaterial, which would further enhance its $T_c$ [2]. Spatial dispersion of a metamaterial is indeed well known to originate from plasmonic effects in its metallic constituents. In a periodic core-shell nanoparticle-based ENZ metamaterial spatial dispersion is defined by the coupling of plasmonic modes of its individual nanoparticles. This coupling enables propagating plasmonic Bloch modes and, hence, nonlocal effects.

Here, we report the first successful realization of such an ENZ core-shell metamaterial superconductor using compressed $Al_2O_3$-coated aluminium nanoparticles, leading to tripling of the metamaterial critical temperature compared to the bulk aluminium. This material is ideal for the proof of principle experiments because the critical temperature of aluminium is quite low ($Tc_{Al}$=1.2K), leading to a very large superconducting coherence length $\xi$=1600 nm [10]. Such a large value of $\xi$ facilitates the metamaterial fabrication requirements while $Al_2O_3$ exhibits very large positive values of dielectric permittivity up to $\varepsilon_{Al2O3}\sim200$ in the THz frequency range [11]. These results provide an explanation for the long known, but not understood, enhancement of the $T_c$ of granular aluminum films [12,13].

The 18 nm diameter Al nanoparticles for these experiments were acquired from the US Research Nanomaterials, Inc. Upon exposure to the ambient conditions a ~ 2 nm thick $Al_2O_3$ shell is known to form on the aluminium nanoparticle surface [14], which is comparable to the 9 nm radius of the original Al nanoparticle. Further aluminium oxidation may also be achieved by heating the nanoparticles in air. The resulting core-shell $Al_2O_3$-Al nanoparticles were compressed into macroscopic, ~1cm diameter, ~0.5 mm thick test pellets using a hydraulic press, as illustrated in the inset in Fig.1.

The IR reflectivity of such core-shell metamaterial samples was measured in the long wavelength IR (LWIR) (2.5-22.5 μm) range using an FTIR spectrometer, and compared with reflectivity spectra of Al and $Al_2O_3$, as shown in Fig. 2. While the reflectivity spectrum of Al is almost flat, the spectrum of $Al_2O_3$ exhibits a very sharp step-like behaviour around 11 μm that is related to the phonon-polariton resonance. This step-like behaviour may be used to characterize the volume fraction of $Al_2O_3$ in the core-shell metamaterial. In the particular case shown in Fig.2, the volume fraction of $Al_2O_3$ in the core-shell metamaterial may be estimated as ~ 39%, which corresponds to $(r_s-r_c)\sim0.18r_c$. At $r_c\sim9$ nm the corresponding thickness of $Al_2O_3$ appears to be $(r_s-r_c)\sim1.6$ nm, which matches expectations based on [14].

The Kramers-Kronig analysis of the FTIR reflectivity spectra of the Al-$Al_2O_3$ sample also allows us to evaluate $\varepsilon_{eff}(0,\omega)$ for the metamaterial in the LWIR spectral range. Plots of the real part of $\varepsilon$ for pure Al and for the Al-$Al_2O_3$ core-shell metamaterial based on the Kramers-Kronig analysis of the data in Fig.2 are plotted in

Fig.3. The plot in Fig.3(a) clearly demonstrates that $\varepsilon_{Al\text{-}Al2O3} \ll \varepsilon_{Al}$ so that the ENZ condition was achieved in the sense that the initial dielectric constant of aluminium was reduced by a factor ~1000. On the other hand, Fig.3(b) demonstrates that the dielectric constant of the Al-Al$_2$O$_3$ core-shell metamaterials remains negative and relatively small above 11 μm. In particular, the large negative contribution to $\varepsilon$ from the aluminium cores is compensated by the large positive contribution from the Al$_2$O$_3$ shells leading to the upturn of $\varepsilon_{Al\text{-}Al2O3}$ that is observed near 20 μm in Fig.3(b) which is caused by large positive value of $\varepsilon_{Al2O3}$ in this spectral range. Note that while both metamaterials shown in Fig.3(b) exhibit much smaller $\varepsilon$ compared to the bulk aluminium, the metamaterial prepared using less oxidized aluminium nanoparticles exhibits considerably larger negative $\varepsilon$. The relatively large noise observed in the calculated plot of $\varepsilon_{Al}$ in Fig.3(a) is due to the fact that the aluminium reflectivity is close to 100% above 7 μm so that the Kramers-Kronig-based numerical analysis of the reflectivity data does not work reliably for pure aluminium samples in this spectral range. Another limitation on the accuracy of the analysis is the use of the finite spectral range (2.5-22.5 μm) of the FTIR spectrometer rather than the infinite one assumed by the rigorous Kramers-Kronig analysis. These limitations notwithstanding, we note that our result for pure aluminium is in good agreement with the tabulated data for $\varepsilon_{Al}$ reported in [15]. Therefore, these results reliably confirm the ENZ character of the core-shell Al-Al$_2$O$_3$ metamaterial. It is also interesting to note that the same FTIR technique applied to the tin-BaTiO$_3$ nanocomposite metamaterials studied in [3] also confirms their expected ENZ character as illustrated in Fig. 4.

The $T_c$ of various Al-Al$_2$O$_3$ core-shell metamaterials was determined via the onset of diamagnetism for samples with different degrees of oxidation using a MPMS

SQUID magnetometer. The zero field cooled (ZFC) magnetization per unit mass versus temperature for several samples with various volume fractions of $Al_2O_3$ is plotted in Fig. 5(a), while the corresponding reflectivity data are shown in Fig.5(b). Even though the lowest achievable temperature with our MPMS SQUID magnetometer was 1.7K, we were able to observe a gradual increase of $T_c$ that correlated with an increase of the $Al_2O_3$ volume fraction as determined by the drop in reflectivity shown in Fig.5(b). The observed increase in $T_c$ also showed good correlation with the results of the Kramers-Kronig analysis shown in Fig.3(b): samples exhibiting smaller negative $\varepsilon$ demonstrated higher $T_c$ increase. The highest onset temperature of the superconducting transition reached 3.9K, which is more than three times as high as the critical temperature of bulk aluminium, $T_{cAl}=1.2K$ [10]. All of the samples exhibited a small positive susceptibility that increased with decreasing temperature, consistent with the presence of small amounts of paramagnetic impurities. The discussed $T_c$ values were determined by the beginning of downturn of M(T), where the diamagnetic superconducting contribution starts to overcome paramagnetic contribution, making this temperature the lower limit of the onset of superconductivity). Further oxidation of aluminium nanoparticles by annealing for 2 hours at 600°C resulted in a $T_c$ less than 1.7K, our lowest achievable temperature. Based on the reflectivity step near 11 μm (see Fig.5(b)), the volume fraction of $Al_2O_3$ in this sample may be estimated as ~ 50%, which corresponds to $(r_s-r_c)$~0.26$r_c$. For $r_c$~9 nm, the corresponding thickness of $Al_2O_3$ was $(r_s-r_c)$~2.4 nm.

Thus, the theoretical prediction of a large increase of $T_c$ in ENZ core-shell metamaterials [2] has been confirmed by direct measurements of $\varepsilon_{eff}(0,\omega)$ of the fabricated metamaterials and the corresponding measurements of the increase of $T_c$. These results strongly suggest that increased aluminium $T_c$'s that were previously observed in thin (~70 nm thickness) granular aluminium films [12,13] were due to changes in the dielectric response function rather than quantum size effects and soft surface phonon modes [13]. As clearly demonstrated by our experimental data and the discussion above, the individual Al nanoparticle size is practically unaffected by oxidation, thus excluding the size effects as an explanation of giant $T_c$ increase in our core-shell metamaterial samples.

**Acknowledgement**

This work was supported in part by NSF grant DMR-1104676.

**Figure Captions**

**Figure 1**. Schematic geometry of the ENZ metamaterial superconductor based on the core-shell nanoparticle geometry. The nanoparticle diameter is d=18 nm. The inset shows typical core-shell metamaterial dimensions.

**Figure 2**. Comparison of the FTIR reflectivity spectrum of a typical core-shell $Al_2O_3$-Al metamaterial sample with reflectivity spectra of bulk Al and $Al_2O_3$ samples. The step in reflectivity around 11 μm may be used to characterize the volume fraction of $Al_2O_3$ in the core-shell metamaterial. The increased noise near 22 μm is related to the IR source cutoff.

**Figure 3**.  The plots of the real part of $\varepsilon$ for pure Al and for the Al-$Al_2O_3$ core-shell metamaterial based on the Kramers-Kronig analysis of the FTIR reflectivity data from Fig.2: (a) Comparison of $\varepsilon$ ' for pure Al and for the Al-$Al_2O_3$ metamaterial clearly indicates that $\varepsilon_{Al-Al2O3} << \varepsilon_{Al}$. (b) Real part of $\varepsilon$ for two different Al-$Al_2O_3$ core-shell metamaterials based on the Kramers-Kronig analysis. While both metamaterials shown in (b) exhibit much smaller $\varepsilon$ compared to the bulk aluminium, the metamaterial prepared using less oxidized aluminium nanoparticles exhibits considerably larger negative $\varepsilon$.

**Figure 4**. The plots of the real part of $\varepsilon$ for pure tin and for the ENZ tin-$BaTiO_3$ nanocomposite metamaterial studied in [3]: (a) Comparison of $\varepsilon$ ' for compressed tin nanoparticles  and for the tin-$BaTiO_3$ nanocomposite metamaterial. (b) Real part of $\varepsilon$ for the tin-$BaTiO_3$ nanocomposite metamaterial.

**Figure 5**. (a) Temperature dependence of zero field cooled magnetization per unit mass for several Al-$Al_2O_3$ core-shell metamaterial samples with increasing degree of oxidation measured in magnetic field of 10 G. The highest onset of superconductivity at ~3.9K is marked by an arrow. This temperature is 3.25 times larger than Tc=1.2K of bulk aluminium.     (b) Corresponding FTIR reflectivity spectra of the core-shell metamaterial samples. Decrease in reflectivity corresponds to decrease of the volume fraction of aluminium.

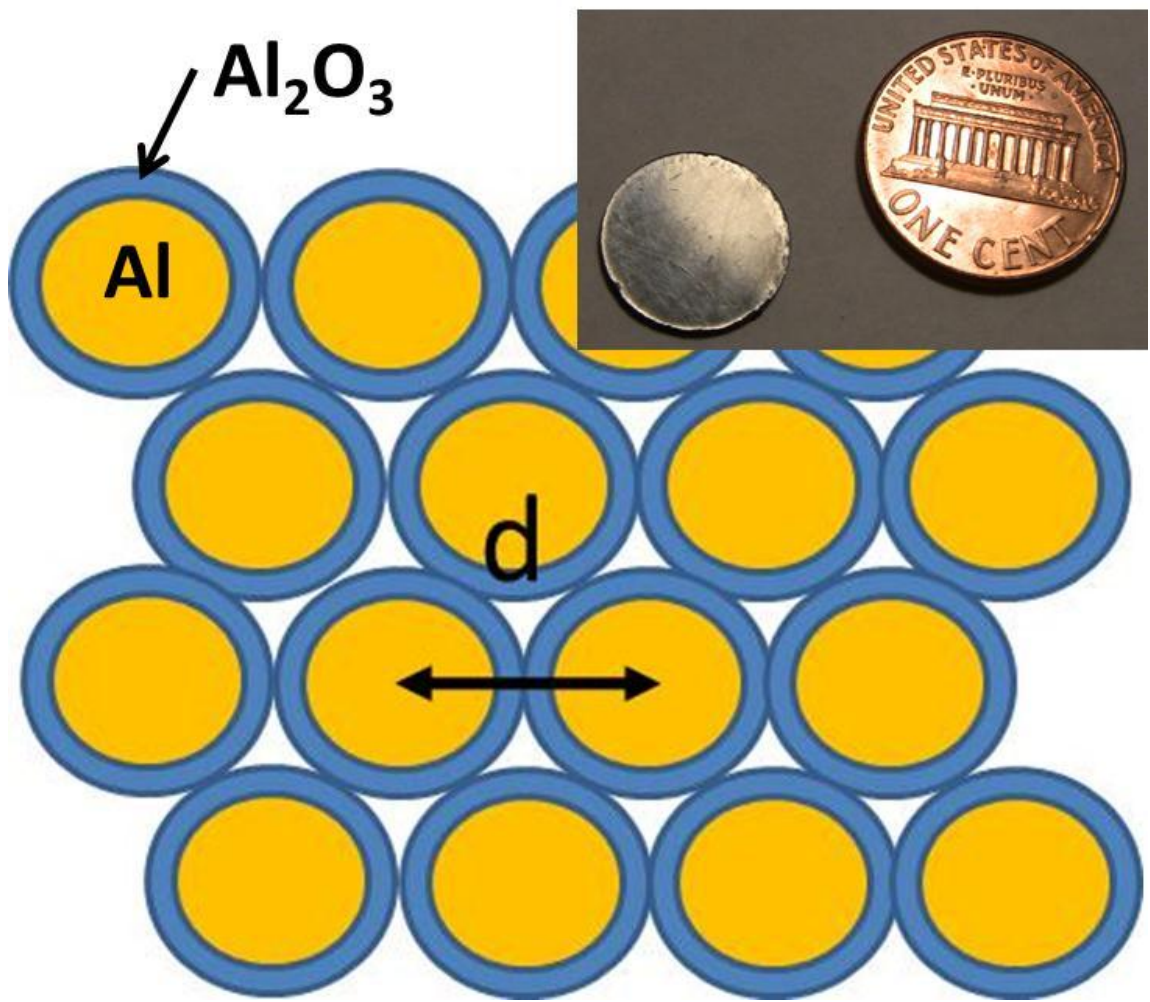

Fig. 1

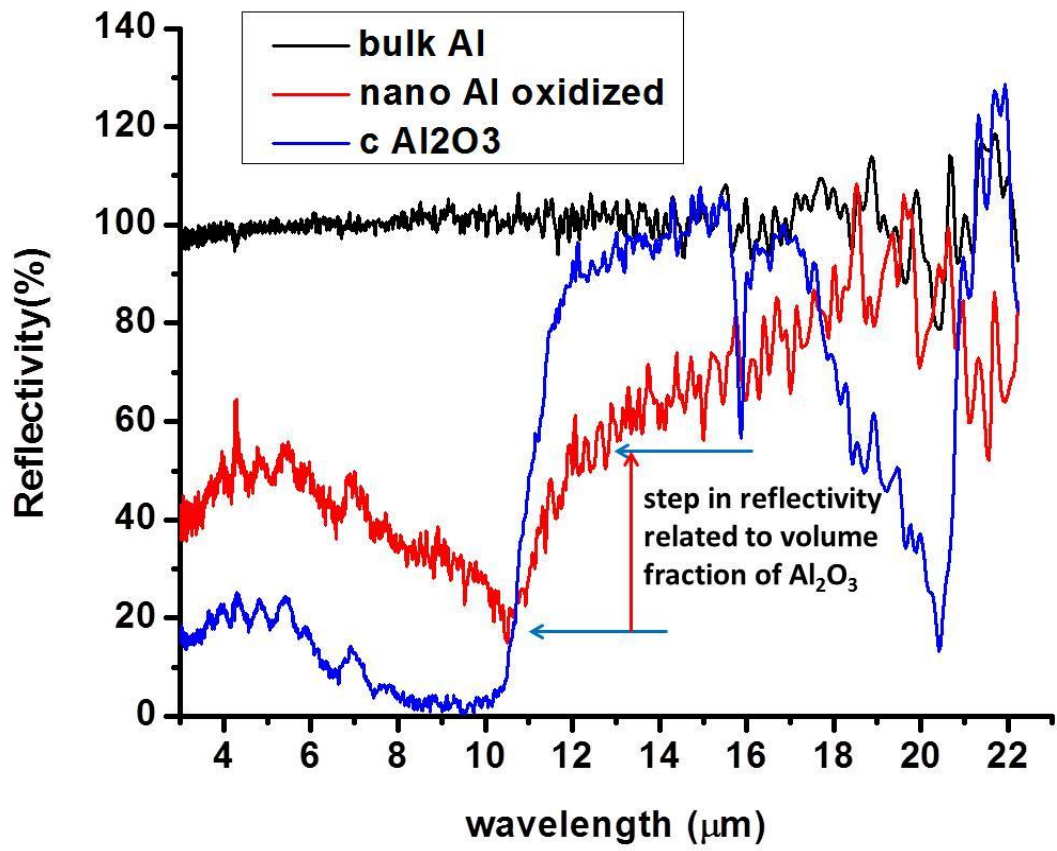

Fig. 2

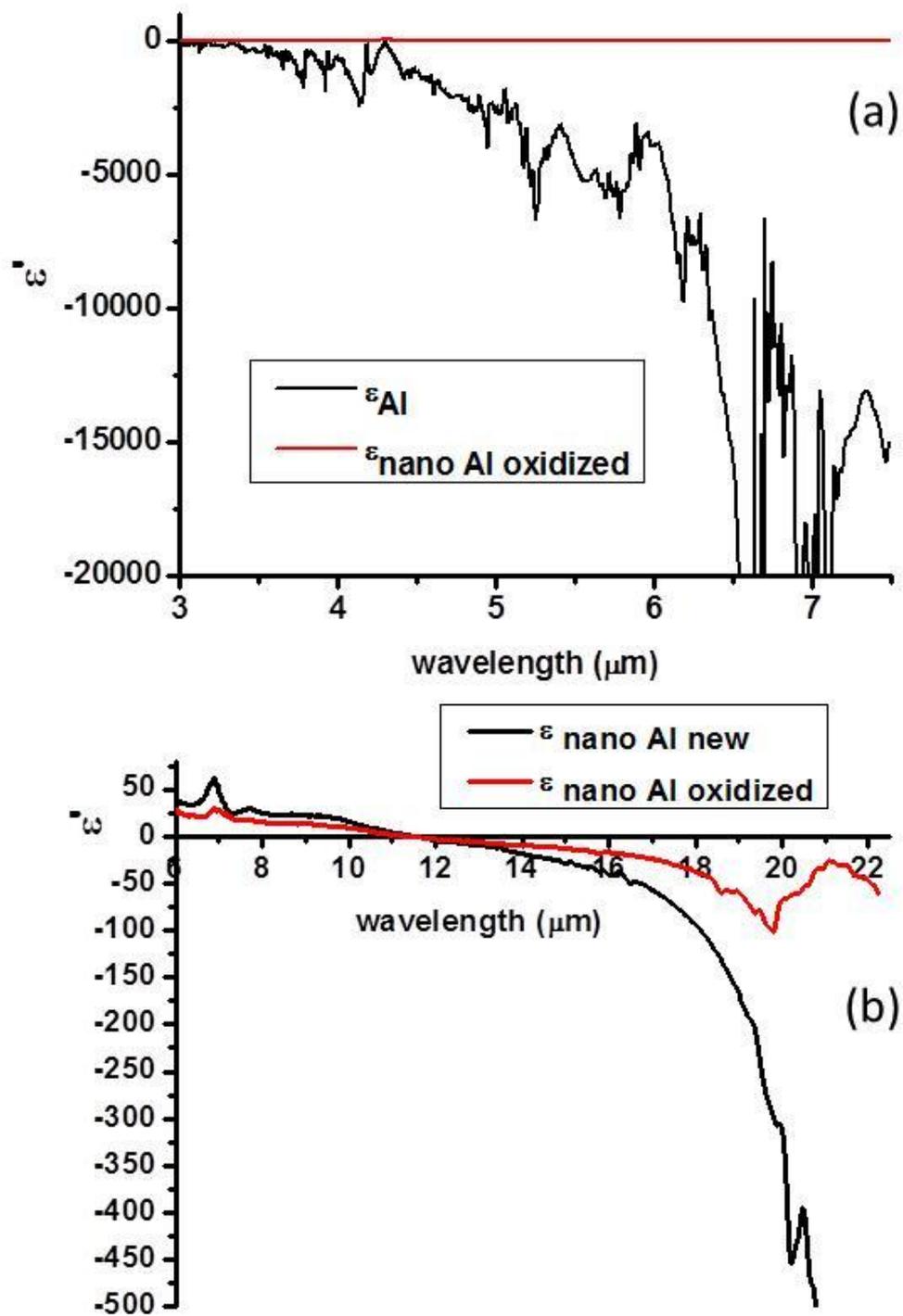

Fig. 3

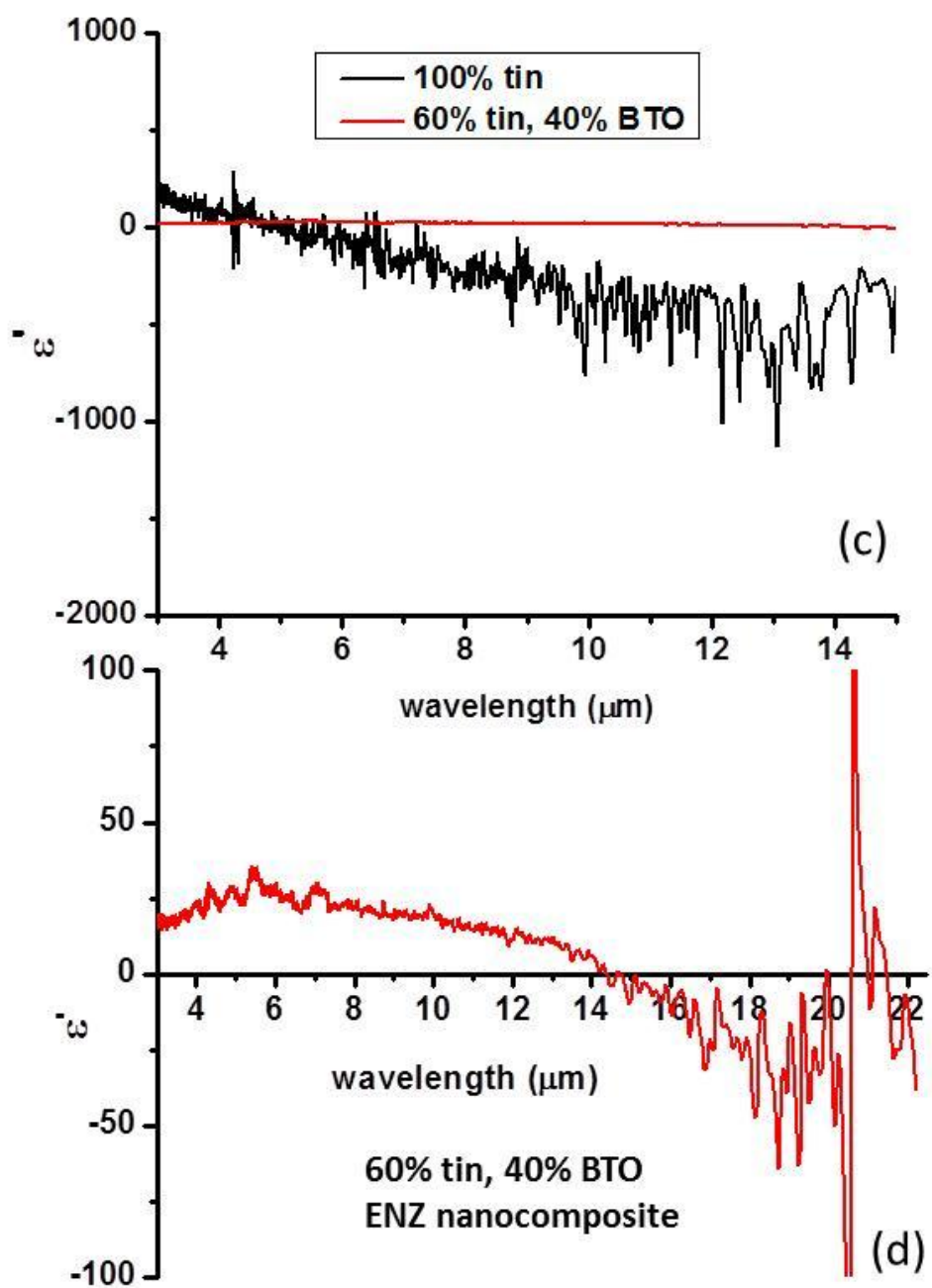

Fig. 4

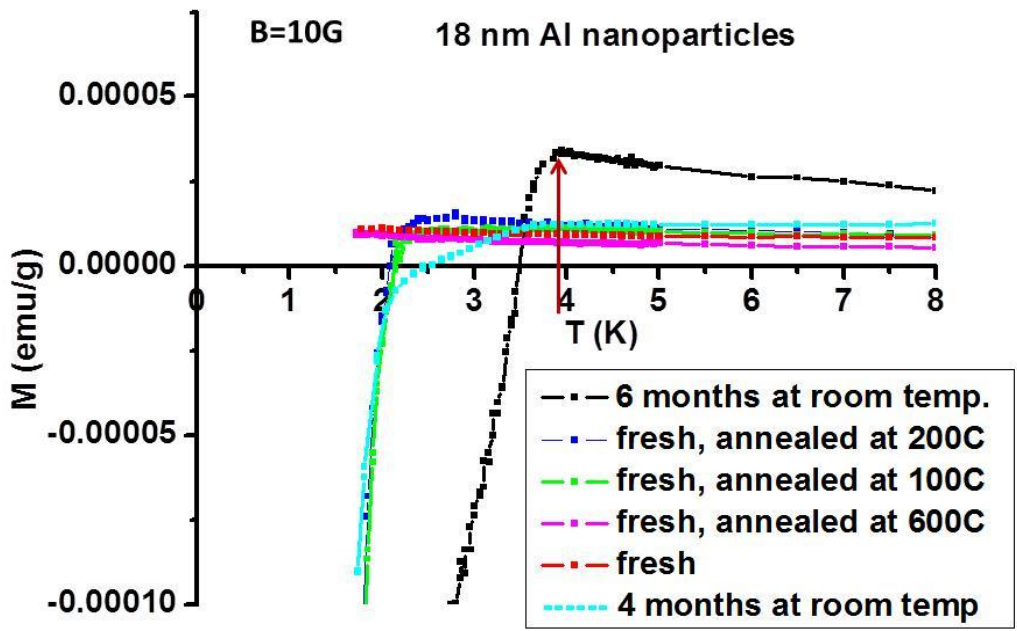

(a)

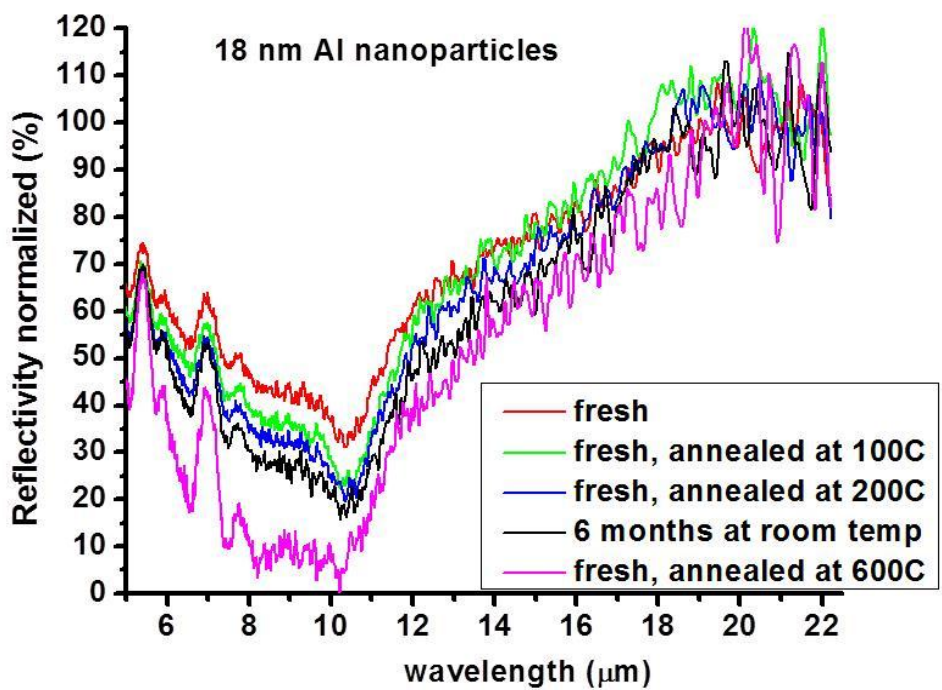

(b)

Fig. 5